# MIMO-Assisted Channel-Based Authentication in Wireless Networks

Liang Xiao, Larry Greenstein, Narayan Mandayam, Wade Trappe
Wireless Information Network Laboratory (WINLAB), Rutgers University
671 Rt. 1 South, North Brunswick, NJ 08902

*Abstract*— Multiple-input multiple-output (MIMO) techniques allow for multiplexing and/or diversity gain, and will be widely deployed in future wireless systems. In this paper, we propose a MIMO-assisted channel-based authentication scheme, exploiting current channel estimation mechanisms in MIMO systems to detect spoofing attacks with very low overhead. In this scheme, the use of multiple antennas provides extra dimensions of channel estimation data, and thus leads to a "security gain" over single-input single-output (SISO) systems. We investigate the security gain of MIMO systems in several system configurations via simulations for a specific real indoor environment using ray-tracing software. We also discuss the effect of increasing the number of transmit and receive antennas on the security gain and contrast that to the diversity/multiplexing gain.

## I. INTRODUCTION

Wireless networks have become pervasive and essential, but most wireless systems lack the ability to reliably identify clients without employing complicated cryptographic tools. This problem introduces a significant threat to the security of wireless networks, since intruders can access wireless networks without a physical connection. One serious consequence is that spoofing attacks (or masquerading attacks), where a malicious device claims to be a specific client by spoofing its MAC address, becomes possible. Spoofing attacks can seriously degrade network performance and facilitate many forms of security weakness, for instance, if attacking control messages/ management frames smartly, the intruder can corrupt services of legal clients [1]–[3].

It is desirable to conduct authentication at the lowest possible layer, and thus a channel-based authentication approach was proposed in [4], exploiting the fact that, in rich multipath environments typical of wireless scenarios, channel responses are *location-specific*. More specifically, channel frequency responses decorrelate from one transmit-receive path to another, if the paths are separated by the order of an RF wavelength or more [5]. Channel-based authentication is able to discriminate among transmitters with low system overhead, since it utilizes existing channel estimation mechanisms.

This prior work [4] on physical layer authentication has focused on single antenna systems. However, with the ability to provide diversity gain and/or multiplexing gain, multiple-input multiple-output (MIMO) techniques will be widely deployed in future wireless networks, e.g. IEEE 802.11 n, to improve traffic capacity and link quality [6]. Therefore, in this paper, we extend the analysis of channel-based authentication to MIMO systems, and investigate the impact of MIMO techniques on the performance of spoofing detection.

We note that the channel-based authentication is used to discriminate among different transmitters, and must be combined with a traditional handshake authentication process to completely identify an entity. We assume that an entity's identity is obtained at the beginning of a transmission using traditional higher layer authentication mechanisms. Channel-based authentication is then used to ensure that all signals in both the handshake process and data transmission are actually from the same transmitter. Thus this may be viewed as a cross-layer design approach to authentication.

We begin the paper by describing the system model in Section II, including the attack model and channel estimation. Then we present our MIMO-assisted channel-based authentication scheme in Section III. In Section IV, we describe the simulation approach and present simulation results. We conclude in Section V with a discussion of the effect of MIMO transmission parameters on the authentication performance. We also contrast the diversity/multiplexing gains with the security gain.

## II. SYSTEM MODEL

### A. Attack Model

Throughout the discussion, we introduce three different parties: Alice, Bob and Eve. As shown in Fig. 1, they are assumed to be located in spatially separated positions. Alice is the legal client with $N_T$ antennas, initiating communication by sending signals to Bob. As the intended receiver, Bob is the legal access point (AP) with $N_R$ antennas. Their nefarious adversary, Eve, will inject undesirable communications into the medium with $N_E$ antennas, in the hopes of impersonating Alice.

In order to obtain the multiplexing gain associated with multiple antennas, the channel state information must be known at receivers [7]. Thus we assume that legal transmitters send non-overlapping pilots from $N_T$ antennas, and Bob uses it to estimate channel responses, for non-security purposes. In the authentication process, Bob tracks the channel responses to discriminate between legitimate signals from Alice and illegitimate signals from Eve.

The authors may be reached at {lxiao, ljg, narayan, trappe}@winlab.rutgers.edu. This research is supported, in part, through grants, CNS-0626439 and CCF-049724, from the National Science Foundation.

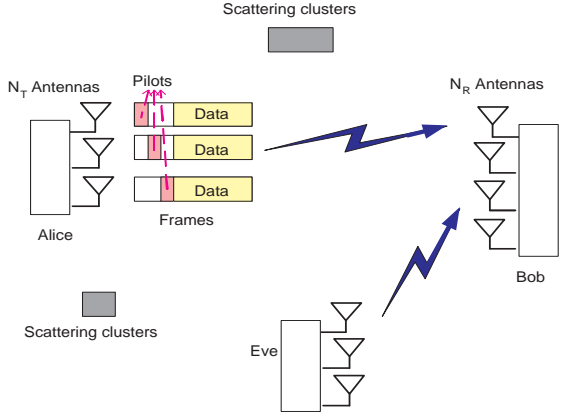

Fig. 1. The adversarial multipath environment involving multiple scattering surfaces. The transmission from Alice with $N_T$ antennas to Bob with $N_R$ antennas, experiences different multipath effects than the transmission by the adversary, Eve. Bob uses pilot symbols to estimate channel responses from the transmitters, and thus discriminate between Alice and Eve.

### B. Channel Estimation Model

A legal transmission from Alice to Bob in Fig. 1 will involve a MIMO system with $N_T$ transmit (Tx) antennas and $N_R$ receive (Rx) antennas. Bob measures and stores channel frequency response samples at $M$ tones, across an overall system bandwidth of $W$, where each subband has bandwidth $b$ ($\leq W/M$), and the center frequency of the system is $f_0$.

We consider channel frequency responses for two frames, which may or may not come from the same transmitter, and denote them by

$$\mathbf{H}_i = [\underline{H}_i(1,1), \underline{H}_i(1,2), \cdots, \underline{H}_i(N_T, N_R)]^T, \quad i = 1, 2, \tag{1}$$

where $\underline{H}_i(j_t, j_r) = [H_{i,1}(j_t, j_r), \cdots, H_{i,M}(j_t, j_r)]^T$, $1 \leq j_t \leq N_T$, $1 \leq j_r \leq N_R$, and $H_{i,m}(j_t, j_r) = H_i(j_t, j_r, f_o + W(m/M - 0.5))$ is the channel response at the $m$-th tone in the $i$-th frame, connecting the $j_t$-th Tx antenna and $j_r$-th Rx antenna. The $N_T N_R M$ elements in $\mathbf{H}_i$ are independent and identically distributed.

In a real receiver, the phase of the local oscillator changes with time, leading to a phase measurement rotation of the underlying channel responses. The phase shifts are the same in channel estimations of $N_R$ antennas, since the antennas are connected to the same receiver oscillator. Considering the phase rotation and receiver thermal noise, we model the estimated channel frequency response as

$$\hat{\mathbf{H}}_i = \mathbf{H}_i e^{j\phi_i} + \mathbf{N}_i, \tag{2}$$

where $\phi_i \in [0, 2\pi)$ denotes the unknown phase measurement rotation, and $\mathbf{N}_i$ is the receiver thermal noise vector with $N_T N_R M$ elements, which are independent and identically distributed complex Gaussian random variables, $CN(0, \sigma^2)$.

The noise variance, $\sigma^2$, is defined as the receiver noise power per tone, $P_N = \kappa T N_F b$, divided by the transmit power per tone per transmit antenna, $P_T/N_T$, i.e.,

$$\sigma^2 = \frac{N_T P_N}{P_T} = \frac{N_T \kappa T N_F b}{P_T}, \tag{3}$$

where $P_T$ is the transmit power per tone, $\kappa T$ is the thermal noise density in mW/Hz, $N_F$ is the receiver noise figure, and $b$ is the measurement noise bandwidth per tone (equals to the subband bandwidth). The signal-to-noise ratio (SNR) in the channel estimation per tone is defined as

$$SNR = \frac{P_T E[||\mathbf{H}_i||_F^2]}{P_N N_T^2 N_R M}, \tag{4}$$

where the expected value is taken over all the channel realizations at locations of interests, and $||A||_F$ denotes the Frobenius norm of the matrix A.

### III. MIMO-ASSISTED AUTHENTICATION

MIMO-assisted channel-based authentication compares channel frequency responses at consecutive frames. Assuming stationary terminals and time-invariant channels, we should report spoofing attacks if channel responses from the same user are significantly different in two frames.

MIMO techniques introduce an extra benefit to spoofing detection. Considering the Alice-Bob-Eve attack model in Fig. 1, if Eve does not know the number of transmit antennas at Alice, $N_T$, she has to predict $N_T$. If Eve has the wrong prediction, or she simply does not have $N_T$ antennas, Bob will foil her with certainty, based on the messed up channel estimation and data decoding results. In other words, Eve has a chance of fooling Bob only if she knows $N_T$ and uses $N_T$ transmit antennas, as is our assumption in the following discussions.

### A. Hypothesis Testing

Assuming Bob obtains channel responses of $\hat{\mathbf{H}}_1$ and $\hat{\mathbf{H}}_2$, respectively, for two frames with the same identity, we build a simple hypothesis test for the purpose of transmitter discrimination. In the null hypothesis, $\mathcal{H}_0$, two estimates are from the same terminal, and thus the claimant is the legal user. Otherwise, Bob accepts the alternative hypothesis, $\mathcal{H}_1$, and claims that a spoofing attack has occurred, i.e., the claimant terminal is no longer the previous one:

$$\mathcal{H}_0: \quad \mathbf{H}_1 = \mathbf{H}_2 \tag{5}$$
$$\mathcal{H}_1: \quad \mathbf{H}_1 \neq \mathbf{H}_2. \tag{6}$$

Since both $\phi_1$ and $\phi_2$ are unknown, Bob chooses the pairwise test statistic as

$$L = \frac{1}{\sigma^2} ||\hat{\mathbf{H}}_1 - \hat{\mathbf{H}}_2 e^{j\phi}||^2, \tag{7}$$

where

$$\phi = \arg\min_x ||\hat{\mathbf{H}}_1 - \hat{\mathbf{H}}_2 e^{jx}|| = Arg(\hat{\mathbf{H}}_1 \hat{\mathbf{H}}_2^H). \tag{8}$$

In the high SNR region, where the proposed scheme must perform, it is easy to show that, under $\mathcal{H}_0$, we have

$$L_{\mathcal{H}_0} \approx \frac{1}{\sigma^2} ||\mathbf{N}_1 - \mathbf{N}_2||^2 \sim \chi_S^2, \tag{9}$$

indicating that $L$ is approximately a Chi-square variable with $S = 2N_T N_R M$ degrees of freedom. Otherwise, when $\mathcal{H}_1$ is true, $L$ is a non-central Chi-square variable, given by

$$L_{\mathcal{H}_1} \approx \frac{1}{\sigma^2} ||\mathbf{H}_1 - \mathbf{H}_2 e^{j\phi} + \mathbf{N}_1 - \mathbf{N}_2||^2 \sim \chi_{S,\mu}^2, \tag{10}$$

where the non-centrality parameter, $\mu$, is written as

$$\mu = \frac{P_T}{P_N N_T} ||\mathbf{H}_1 - \mathbf{H}_2 e^{jArg(\mathbf{H}_1 \mathbf{H}_2^H)}||^2. \quad (11)$$

For fixed $P_T$, the dimension of $\mathbf{H}_i$ is proportional to $MN_R$, and thus $\mu$ rises with both $N_R$ and $M$. On the other hand, the impact of $N_T$ is more complex, depending on the specific value of $\mathbf{H}_1$, $\mathbf{H}_2$, and $P_T$.

The rejection region of $\mathcal{H}_0$ is defined as $L \leq k$, where $k$ is the test threshold, which is selected according to an appropriate performance target.

### B. Performance Criteria

Given a building environment and terminal locations, we derive the performance of MIMO-based spoofing detection, averaged over all realizations of receiver thermal noise. From Eq. (9), we can write the "false alarm rate" (or Type I error) for a given $k$ as

$$\alpha = Pr(L > k|\mathcal{H}_0) = 1 - F_{\chi_S^2}(k), \quad (12)$$

where $F_X(\cdot)$ is the CDF of the random variable $X$. Similarly, from Eq. (10), the "miss detection rate" (or Type II error) for given $k$ is given by

$$\beta = Pr(L \leq k|\mathcal{H}_1) = F_{\chi_{S,\mu}^2}(k), \quad (13)$$

indicating that $\alpha$ rises with $k$, while $\beta$ decreases with it. By Eq. (12) and (13), we have the miss rate for given false alarm rate as

$$\beta(\alpha) = F_{\chi_{S,\mu}^2}(F_{\chi_S^2}^{-1}(1-\alpha)), \quad (14)$$

where $F_X^{-1}(\cdot)$ is the inverse function of $F_X(\cdot)$. From Eq. (11) and (14), we see the miss rate decreases with $P_T$, since higher transmit power allows for more accurate channel estimation.

We will investigate the security gain of MIMO techniques in our channel-based authentication scheme. For given $\alpha$, it is defined as the relative decrease of $\beta(\alpha)$, if replacing single antenna systems with multiple antenna systems, i.e.,

$$G = \frac{\beta_{SISO}(\alpha) - \beta_{MIMO}(\alpha)}{\beta_{MIMO}(\alpha)}, \quad (15)$$

where $\beta_{SISO}$ and $\beta_{MIMO}$ are the miss rates in the single antenna systems and multiple antenna systems, respectively.

### C. Performance Discussion

The use of multiple antennas has a two-fold impact: it improves security performance by increasing the frequency sample size from $2M$ to $2MN_TN_R$. On the other hand, the use of multiple transmit antennas reduces the transmit power per antenna, leading to performance loss of some degree.

Note that the frequency sample size, $M \in [1, M_s]$, is selected for security purposes, where $M_s$ ($\geq M$), the total number of subbands, is determined by non-security issues such as data decoding accuracy. The average transmit power per tone is determined by $M_s$, with $P_T = P_{total}/M_s$, where $P_{total}$ is the total system transmit power. Hence, $P_T$ is independent of any other parameters mentioned, and we assume constant $P_T$ in the comparison of system configurations.

In wideband systems, $b$ is fixed and the detection performance improves with $W$, since channel responses decorrelate more rapidly in space with higher system bandwidth. From (3), (11), and (14), we see that $\beta$ increases with $b$, since the power of measurement noise is proportional to $b$. As will be shown later, the optimal choice for wideband systems is to set $M = M_s$.

In narrowband systems, however, since $W < B_c$, where $B_c$ is the channel coherence bandwidth, we set $M = 1$ and $W = b$. As a result, the detection performance improves as system bandwidth $W = b$ decreases, as can be inferred from Eq. (3), (11), and (14).

## IV. SIMULATION AND NUMERICAL RESULTS

### A. Simulation Method

The WiSE tool, a ray-tracing software package developed by Bell Laboratories [8], was used to model not only typical channel responses, but the spatial variability of these responses. One input to WiSE is the 3-dimensional plan of a specific building, including walls, floors, ceilings and their material properties (e.g., dielectric coefficient and conductivity). With this information, WiSE calculates the rays at any receiver from any transmitter, including their amplitudes, phases and delays. From this, it is straightforward to construct the transmit-receive frequency response over any specified interval.

We have done this for a typical office building, for which a top view of the first floor is shown in Fig. 2. This floor of this building is 120 meters long, 14 meters wide and 4 meters high. For our numerical experiment, we placed the access point (AP) in the hallway at [45.6, 6.2, 3.0] m. For the positions of transmitters, we considered a 12 m × 67 m area, shown as outlined with a dashed line in the figure. We assumed all transmitters are at a height of 2 m, being anywhere on a uniform horizontal grid of 405 points with 1.5-meter spacing.

We randomly chose 2 points within the 12 m × 67 m area as the legal and spoofing nodes. For each scenario, (1) WiSE was used to generate channel impulse responses for the 2 nodes; and (2) the hypothesis test described above was used to compute $\beta$, for given $\alpha$, by Eq. (14). We repeated the experiment $405 \times 404/2 = 81810$ times, and computed the average miss rate, for each system configuration.

### B. Simulation Results

In the simulations, we consider MIMO, single-input multiple-output (SIMO), multiple-input single-output (MISO), and single-input single-output (SISO) systems, with seperation of two neighboring antennas of 3 cm (i.e., half wavelength), $\alpha = 0.01$, $f_0 = 5$ GHz, $N_F = 10$, $b = 0.25$ MHz, and $P_T \in \{0.1, 1, 10\}$ mW, if not specified otherwise. The per tone SNR ranges from -16.5 dB to 53.6 dB, with a median value of 16 dB, using transmit power per tone $P_T = 0.1$ mW, $b = 0.25$ MHz, and $N_T = N_R = 1$.

Figure 3 shows that the average miss rate decreases with the frequency sample size, $M$, with $W = 20$ MHz, indicating that we should use all of the channel estimation data and set $M = M_s$. In addition, it can be seen that the security gain of MIMO, defined by Eq. (15), decreases with $M$, when $P_T >$ 0.1 mW. For instance, $G(P_T = 1 \text{ mW}, M = 1) = (0.09 - 0.01)/0.01 = 8$, is greater than $G(P_T = 1 \text{ mW}, M = 10) = 1.7$. If using high power and small $M$ (e.g., $M = 1$), the SISO system has accurate but insufficient channel response samples. Thus the additional dimensions of channel samples in MIMO systems allow for much better performance. On the contrary, if using high $P_T$ and large $M$, the performance of SISO systems is too good to be significantly improved.

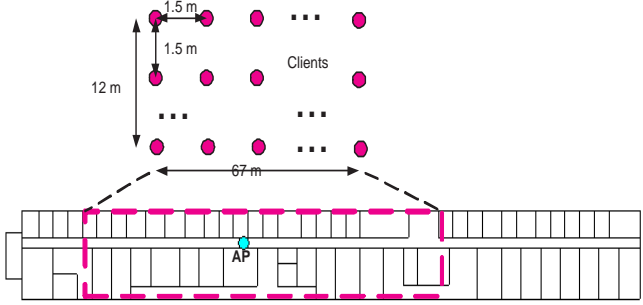

Fig. 2. System topology assumed in the simulations. The receiver is located at [45.6, 6.2, 3.0] m in a 120 m × 14 m × 4 m office building. The antenna distance is half wavelength (3 cm). All transmitters, including both legal transmitters and spoofing nodes, are located on dense grids at a height of 2 m. The total number of samples in the grids is 405.

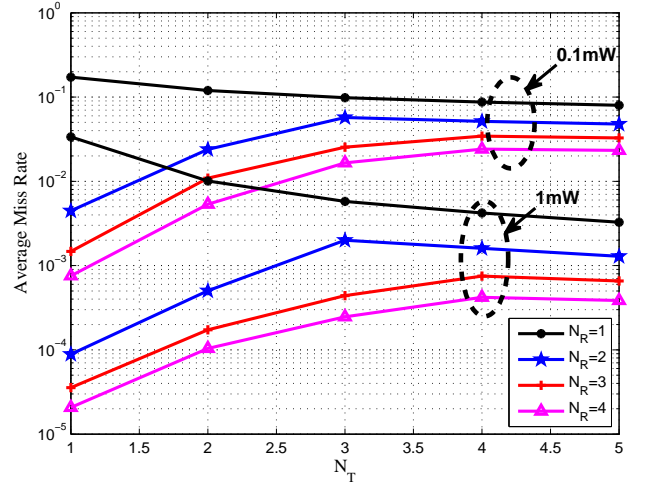

Fig. 4. Average miss rate of spoofing detection for various configuration of $N_T$ and $N_R$, with $\alpha = 0.01$, $M = 3$, $P_T \in \{0.1, 1\}$ mW, $b = 0.25$ MHz, and $W = 2$ MHz.

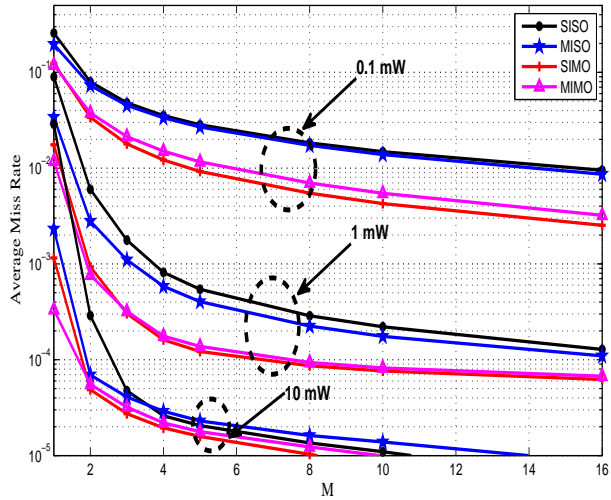

Fig. 3. Average miss rate of spoofing detection in wideband systems, in SISO, $2 \times 1$ MISO, $1 \times 2$ SIMO, and $2 \times 2$ MIMO systems, respectively, with $\alpha = 0.01$, $M = 5$, $b = 0.25$ MHz, $W = 20$ MHz, and $P_T \in \{0.1, 1, 10\}$ mW.

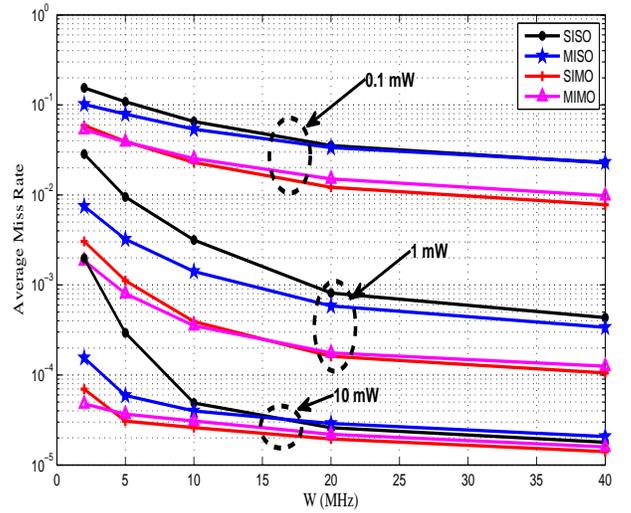

Fig. 5. Average miss rate of spoofing detection in wideband systems, given false alarm rate of 0.01, in SISO, $2 \times 1$ MISO, $1 \times 2$ SIMO, and $2 \times 2$ MIMO systems, respectively, with $\alpha = 0.01$, $M = 4$, $b = 0.25$ MHz, and $P_T \in \{0.1, 1, 10\}$ mW.

We can also see that the security gain slightly rises with $M$, when $P_T$ is as low as 0.1 mW, e.g., $G(P_T = 0.1 \text{ mW}, M = 1) < G(P_T = 0.1 \text{ mW}, M = 10)$. This observation arises, because when the channel estimation is not accurate due to low SNR, the systems need much more data to make a right decision.

Similarly, the impact of $P_T$ on the MIMO security gain also depends on the value of $M$: The gain rises with $P_T$, under small $M$, e.g., $G(P_T = 10 \text{ mW}, M = 1) > G(P_T = 0.1 \text{ mW}, M = 1)$. Otherwise, under large $M$, the security gain decreases with $P_T$, e.g., $G(P_T = 10 \text{ mW}, M = 10) < G(P_T = 0.1 \text{ mW}, M = 10)$.

Next, Fig. 4 indicates that the miss rate decreases with $N_R$, and the security gain of $N_R$ decreases with $N_R$. On the other hand, the impact of multiple ($N_T$) transmit antennas on the authentication performance is determined by parameters like $P_T$, $M$, and $N_R$, since the use of more transmit antennas reduces the transmit power per antenna, while providing additional channel estimation samples. For instance, with $P_T = \{0.1$ mW, $1$ mW$\}$ and $M = 3$, the miss rate decreases with $N_T$, under $N_R = 1$, while it rises with $N_T$, under $N_R > 1$.

As discussed in Section III-C, Fig. 5 shows that the miss rate decreases with system bandwidth, $W$, since the $M=4$ channel samples are less correlated with wider bandwidth. On the other hand, the MIMO security gain decreases with $W$, as the miss rate in SISO systems decreases more rapidly with $W$ than that in MIMO systems. It is also shown that SIMO is better than MIMO, under large $W$.

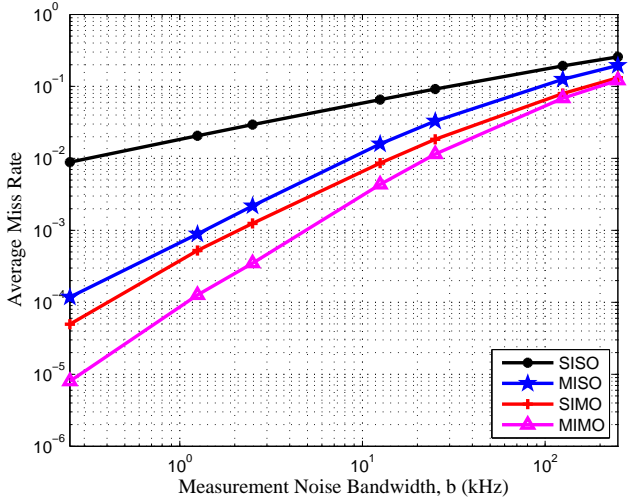

Fig. 6. Average miss rate of spoofing detection in narrowband systems, given false alarm rate of 0.01, in SISO, $2\times 1$ MISO, $1\times 2$ SIMO, and $2\times 2$ MIMO systems, respectively, with $\alpha=0.01$, $M=1$, $P_T=0.1$ mW, and $b=W$.

Finally, the detection performance in narrowband systems is presented in Fig. 6, with $b$ ranging between 250 Hz and 250 kHz. Since a larger noise bandwidth decreases SNR, it raises the miss rate and reduces the MIMO security gain.

## V. Summaries & Discussion

We have proposed a MIMO-assisted channel-based authentication scheme, exploiting the spatial decorrelation property of the wireless medium to detect spoofing attacks. We presented the average miss detection rate, for a given false alarm rate of 0.01, and evaluated the security gain (defined as the improvement in authentication performance over SISO systems, Eq. (15)) for different MIMO transmission parameters. We had the following observations:

- The MIMO security gain decreases with the system bandwidth ($W$), because the SISO system provides sufficient decorrelation at high bandwidth, making resolution of Alice and Eve better.
- The MIMO security gain decreases with the noise bandwidth ($b$) in narrowband systems, since the noise power is larger there by affecting the estimation of MIMO channel parameters.
- The MIMO security gain decreases with the frequency sample size ($M$), if the transmit power ($P_T$) is as large as 1 mW. If using high power and small $M$, the SISO system has accurate but insufficient channel response samples. Thus the additional dimensions of channel samples in MIMO systems allow for much better performance. On the contrary, if using high $P_T$ and large $M$, the performance of SISO systems is too good to be significantly improved.
- On the other hand, the MIMO security gain slightly rises with $M$, if $P_T$ is as small as 0.1 mW. This is because when the channel estimation is not accurate due to low SNR, the systems need much more data to make a right decision.
- Similarly, the MIMO security gain rises with $P_T$, under small $M$ (e.g., $M=1$). Otherwise, it decreases with $P_T$, under large $M$ (e.g., $M=10$).

We can also compare the security gain with the MIMO diversity gain, as a function of the number of transmit and receive antennas. It is well known that the diversity gain rises with both the number of transmit antennas and the number of receive antennas. We have found that

- The use of multiple (i.e., $N_R > 1$) receive antennas improves the detection of spoofing attacks. This is a case where both the security gain and the diversity gain increase due to additional receive antennas.
- On the other hand, the security gain by using multiple (i.e., $N_T > 1$) transmit antennas may be positive or negative, based on the value of $P_T$, $M$, and $N_R$, since the transmit power per antenna decreases with $N_T$, while more transmit antennas provide extra channel estimation samples. This is a case where the security gain sometimes decreases but the diversity gain always rises due to additional transmit antennas.

Thus the MIMO-assisted channel-based authentication schemes provide a wide range of parameter choices and performance tradeoffs that have to be considered in the context of both security gains and MIMO performance gains.